\documentclass[showpacs,preprintnumbers,amsmath,amssymb]{revtex4}
\usepackage{amsmath,amssymb,graphics,epsfig,subfigure}
\usepackage{color}

\newcommand{\tcb}{\textcolor{blue}}

\begin{document}
\renewcommand{\baselinestretch}{1.3}

\title{Intriguing microstructures of five-dimensional neutral Gauss-Bonnet AdS black hole}

\author{Shao-Wen Wei \footnote{weishw@lzu.edu.cn},
  Yu-Xiao Liu \footnote{liuyx@lzu.edu.cn}}

\affiliation{Institute of Theoretical Physics $\&$ Research Center of Gravitation, Lanzhou University, Lanzhou 730000, People's Republic of China}

\begin{abstract}
In this paper, we analytically study the phase structure and construct the Ruppeiner geometry in the extended phase space for the five-dimensional neutral Gauss-Bonnet AdS black hole. Through calculating the scalar curvature of the Ruppeiner geometry and combining the phase transition, we show that the attractive interaction is dominant in the microstructure of the black hole system. More significantly, there is an intriguing property that the normalized scalar curvature has the same expression for the saturated small and large black hole curves. This implies that although the microstructure is different before and after the small-large black hole phase transition, the interaction between the microscopic constituents keeps unchanged. These results are quite valuable on further understanding the microstructure of the AdS black hole in modified gravity.
\end{abstract}



\maketitle

\section{Introduction}

The study of black hole thermodynamics and phase transition has been one of the  increasingly active areas among the last couple of decades. In particular, black hole chemistry attracts much more attention. The cosmological constant was treated as a thermodynamic pressure and its conjugate quantity as volume in the extended phase space \cite{Kastor}. Rich phase transitions and phase structures were observed, for examples, the small-large black hole phase transition, reentrant phase transition, triple point, $\lambda$-line phase transition \cite{Kubiznak,Altamirano,AltamiranoKubiznak,Altamirano3,Dolan,Liu,Wei2,Frassino,
Cai,XuZhao,Kostouki,Hennigar,Hennigar2,Tjoa2,Ruihong}, for a recent review, see Refs. \cite{Altamiranoa,Teob}.

It is extensively shown that the small-large black hole phase transition is extremely similar to the liquid-gas phase transition of Van der Waals (VdW) fluids. As we know, VdW fluids consist of micromolecules with finite size and interaction between them. Therefore, a natural question is what is the microstructure of the black holes. In order to examine it, we proposed that an AdS black hole is also constructed by some unknown micromolecules \cite{Weiw}. Combining with the black hole phase transition, we studied the properties of the black hole microstructures. It was also shown that the interaction between these micromolecules can be tested by the Ruppeiner geometry. This meaningful approach has been generalized to other black holes in AdS space \cite{Dehyadegari,Moumni,Deng,Sheykhi,Miao,Miao2,Miao3,Li,Chen,Guo,Du,Xuz}.

In Ref. \cite{Weiw}, we chose the entropy as the thermodynamic potential and took the black hole mass and the pressure as the fluctuation coordinates. Then we constructed the Ruppeiner geometry and calculated the corresponding scalar curvature. Although some interesting properties were reflected by the scalar curvature, it does not go to negative infinity at the critical point, where a second order phase transition takes place. This particular property is not consistent with the original idea of the Ruppeiner geometry \cite{Ruppeiner}. This inconsistency is generally neglected in previous study.

For the purpose, we further investigated this issue \cite{LiuLiu,WeiWeiWei}, recently. Starting with the Boltzmann's entropy formula, the general Ruppeiner geometry was constructed. Then taking the temperature and volume as the fluctuation coordinates, a universal Ruppeiner metric was worked out and the corresponding scalar curvature was calculated. For a VdW fluid, the scalar curvature indicates that the attractive interactions dominate amongst the fluid microstructures. While for the charged AdS black hole, we found that the Ruppeiner geometry will become problematic due to the vanishing heat capacity. In order to cure this problem, we introduced a new quantity, the normalized scalar curvature, for the charged AdS black hole \cite{LiuLiu}. This treatment can be understood as that the black hole microstructure not completely depends on the heat capacity. For example, VdW fluid has a constant heat capacity $3k_{B}/2$. And it almost has no influence on its microstructure. Employing this new scalar curvature, we observed that the repulsive interaction can exist for the small black hole of high temperature. This novel property shows that, even they share the similar phase transition and critical phenomena, there still exists a large difference between the microstructures of a black hole and VdW fluid. This property was also observed for the higher-dimensional charged AdS black holes \cite{WeiWeiWei}.

Furthermore, we also examined the critical phenomena of the normalized scalar curvature \cite{WeiWeiWei}. At the critical point, the normalized scalar curvature $R_{N}$ goes to negative infinity, which indicates that the correlation length tends to infinity. It was also found that the normalized scalar curvature has a universal exponent $1/2$. Another universal dimensionless parameter is $R_{N}(1-\tilde{T})^{2}$, where $\tilde{T}$ is the reduced temperature. For the VdW fluid, numerical result shows that it is $-1/8$ near the critical point. For the four-dimensional charged AdS black hole, the analytical result confirms this constant. While for higher-dimensional charged AdS black holes, the values are more negative than $-1/8$. Therefore, the observation of the same of the critical phenomena and the universal dimensionless parameter further confirms that the normalized scalar curvature $R_{N}$ can be used to test the black hole microstructures. We also expect to uncover more information of the black hole microstructures.

Since the analytical results can give us the exact information of the black hole microstructures, we would like to further study them. As we know, the coexistence curve of the five-dimensional neutral Gauss-Bonnet (GB) AdS has an analytical form \cite{Mol}. This gives us a good chance to exactly investigate it. The result will provide us a possible way to understand the microstructures of the black hole in this modified gravity.

This work is organized as follows. In Sec. \ref{iner}, we briefly introduce the Ruppeiner geometry. Thermodynamics and phase diagrams of the GB-AdS black hole are discussed in Sec. \ref{Therm}. Then we apply the geometrical method to the black hole in Sec. \ref{rgsc}. The black hole microstructures are studied in detail. Furthermore, the critical phenomena of the normalized scalar curvature are investigated. Finally, the conclusions and discussions are given in Sec. \ref{Conclusion}.

\section{Ruppeiner geometry}
\label{iner}

In this section, we would like to briefly review the Ruppeiner geometry \cite{Ruppeiner}.

Let us first consider a thermodynamic system, which contains two parts, one is a small subsystem S, and the other is its environment E. Thus the total entropy of the system with two independent thermodynamic variables $x^0$ and $x^1$ can be written as
\begin{equation}
 S(x^{0},x^{1})=S_{\rm S}(x^{0},x^{1})+S_{\rm E}(x^{0},x^{1}),
\end{equation}
where we require $S_{\rm S}\ll S_{\rm E}\sim S$. Here we require that the variables $x^0$ and $x^1$ are extensive quantities of the system. Near the local maximum of the entropy at $x^{\mu} = x^{\mu}_0$, the total entropy can be expanded in the following form
\begin{eqnarray}
 S&=&S_{0} + \left. \frac{\partial S_{\rm S}}{\partial x^{\mu}}  \right|_{x^{\mu}_0} \Delta x^{\mu}_{\rm S}
      + \left.\frac{\partial S_{\rm E}}{\partial x^{\mu}}   \right|_{x^{\mu}_0}   \Delta x^{\mu}_{\rm E}\nonumber\\
 &&   + \left. \frac{1}{2}\frac{\partial^{2}S_{\rm S}}{\partial x^{\mu}\partial x^{\nu}}
        \right|_{x^{\mu}_0}  \Delta x^{\mu}_{\rm S}\Delta x^{\nu}_{\rm S}
      + \left. \frac{1}{2}\frac{\partial^{2}S_{\rm E}}{\partial x^{\mu}\partial x^{\nu}}
        \right|_{x^{\mu}_0}  \Delta x^{\mu}_{\rm E}\Delta x^{\nu}_{\rm E}
   +\cdots, \quad (\mu, \nu=0,1),
 \end{eqnarray}
where $S_{0}$ is the local maximum of the entropy $S(x^{\mu}_0)$. Considering that the fluctuating parameters are additive, i.e., $x_{S}^{\mu}+x_{E}^{\mu}=x_{total}^{\mu}=constant$, we can get $\left.\frac{\partial S_{\rm S}}{\partial x^{\mu}} \right|_{0} \Delta x^{\mu}_{\rm S}
   + \left.\frac{\partial S_{\rm E}}{\partial x^{\mu}} \right|_{0} \Delta x^{\mu}_{\rm E}=0$, and then, we arrive
\begin{equation}
 \Delta S=\left. \frac{1}{2}\frac{\partial^{2}S_{\rm S}}{\partial x^{\mu} \partial x^{\nu}}
              \right|_{0} \Delta x^{\mu}_{\rm S} \Delta x^{\nu}_{\rm S}
          +   \left. \frac{1}{2}\frac{\partial^{2}S_{\rm E}}{\partial x^{\mu} \partial x^{\nu}}
              \right|_{0}  \Delta x^{\mu}_{\rm E} \Delta x^{\nu}_{\rm E}
          +   \cdots\;.
 \end{equation}
\tcb{The second term can be ignored because the fluctuations are bigger in smaller systems.} Then, the probability of finding the system in the internals $x_0 + \Delta x_0$ and $x_1 + \Delta x_1$ will be of the following form
\begin{eqnarray}
 P(x^{0},x^{1}) \propto e^{-\frac{1}{2}\Delta l^{2}},
 \end{eqnarray}
 where
 \begin{eqnarray}
  \Delta l^{2}&=&-\frac{1}{k_{\rm B}}g_{\mu\nu} \Delta x^{\mu}\Delta x^{\nu},\label{Ds}\\
 g_{\mu\nu}&=& \left.\frac{\partial^{2}S_{\rm S}}{\partial x^{\mu}\partial x^{\nu}} \right|_{0} .\label{gmunu}
 \end{eqnarray}
According to the thermodynamic information geometry, $\Delta l^{2}$ measures the distance between two neighboring fluctuation states. Different from the Weinhold geometry \cite{Weinhold}, the thermodynamic potential of the Ruppeiner geometry is the entropy rather the internal energy of the system. Given the metric (\ref{gmunu}), we can calculate the scalar curvature for the geometry following GR approach. Then according to the interpretation that positive (negative) Ruppeiner scalar curvature indicates the repulsive (attractive) interaction, we are allowed to test the property of the system microstructures.

\tcb{The line element (\ref{Ds}) can be reexpressed as another form by the coordinate transformation, and under which the line element is required to be invariant. Here it is worthwhile pointing out that the coordinate of the line element $x^{\mu}$ given in (\ref{gmunu}) is extensive quantity. However, after some coordinate transformations, $x^{\mu}$ can be the intensive quantity. In the original paper \cite{Ruppeiner}, the coordinates are changed to the temperature $T$ and the fluid density $\rho$ for the VdW fluid by using the first law of the thermodynamics. They observed the expected property of the fluid microstructures and the critical phenomena of the scalar curvature. The similar approach was also extended to other fluid systems. The general result confirms that negative (positive) scalar curvature indicates attractive (repulsive) interaction of the fluid. In order to generalize this approach to the black hole case, we adopt the same approach. However for the black hole cases, their density has no simple relation $\rho=1/V$. So we will take the temperature $T$ and thermodynamic volume $V$ as the fluctuation coordinates.} 

\tcb{Here, we take the internal energy $U$ and $V$ as the fluctuation coordinates. Then by using the first law of the black hole thermodynamics, the line element can be reexpressed as the following form}
\begin{equation}
 d l^{2}=-\frac{1}{T}\left(\frac{\partial^{2}F}{\partial T^{2}}\right)_{V}d T^{2}
   +\frac{1}{T}\left(\frac{\partial^{2}F}{\partial V^{2}}\right)_{T}d V^{2}. \label{ddl2}
\end{equation}
Here $F=U-TS$ is the free energy, and it satisfies the differential law $dF=-SdT-PdV$. With the help of the heat capacity at constant volume $C_{V}=T(\partial_{T}S)_{V}$, the line element can be further expressed as
\begin{equation}
 d l^{2}=\frac{C_{V}}{T^{2}}d T^{2}-\frac{(\partial_{V}P)_{T}}{T}d V^{2}.\label{xxy}
\end{equation}
This diagonalized metric has the following scalar curvature \cite{WeiWeiWei}
\begin{eqnarray}
 R&=&\frac{1}{2C_{V}^{2}(\partial_{V}P)^{2}}
 \bigg\{
 T(\partial_{V}P)\bigg[(\partial_{T}C_{V})(\partial_{V}P-T\partial_{T,V}P)
 +(\partial_{V}C_{V})^{2}\bigg]\nonumber\\
 &+&C_{V}\bigg[(\partial_{V}P)^{2}+T\left((\partial_{V}C_{V})(\partial_{V,V}P)
 -T(\partial_{T,V}P)^{2}\right)
 +2T(\partial_{V}P)(T(\partial_{T,T,V}P)-(\partial_{V,V}C_{V}))\bigg]
 \bigg\}.\nonumber\\\label{RR}
\end{eqnarray}
Adopting this formula, one can obtain the scalar curvature for the Ruppeiner geometry. Then the microscopic property could be revealed from the scalar curvature.

\section{Thermodynamics and phase transition}
\label{Therm}

\begin{figure}
\center{\subfigure[]{\label{PTphase}
\includegraphics[width=7cm]{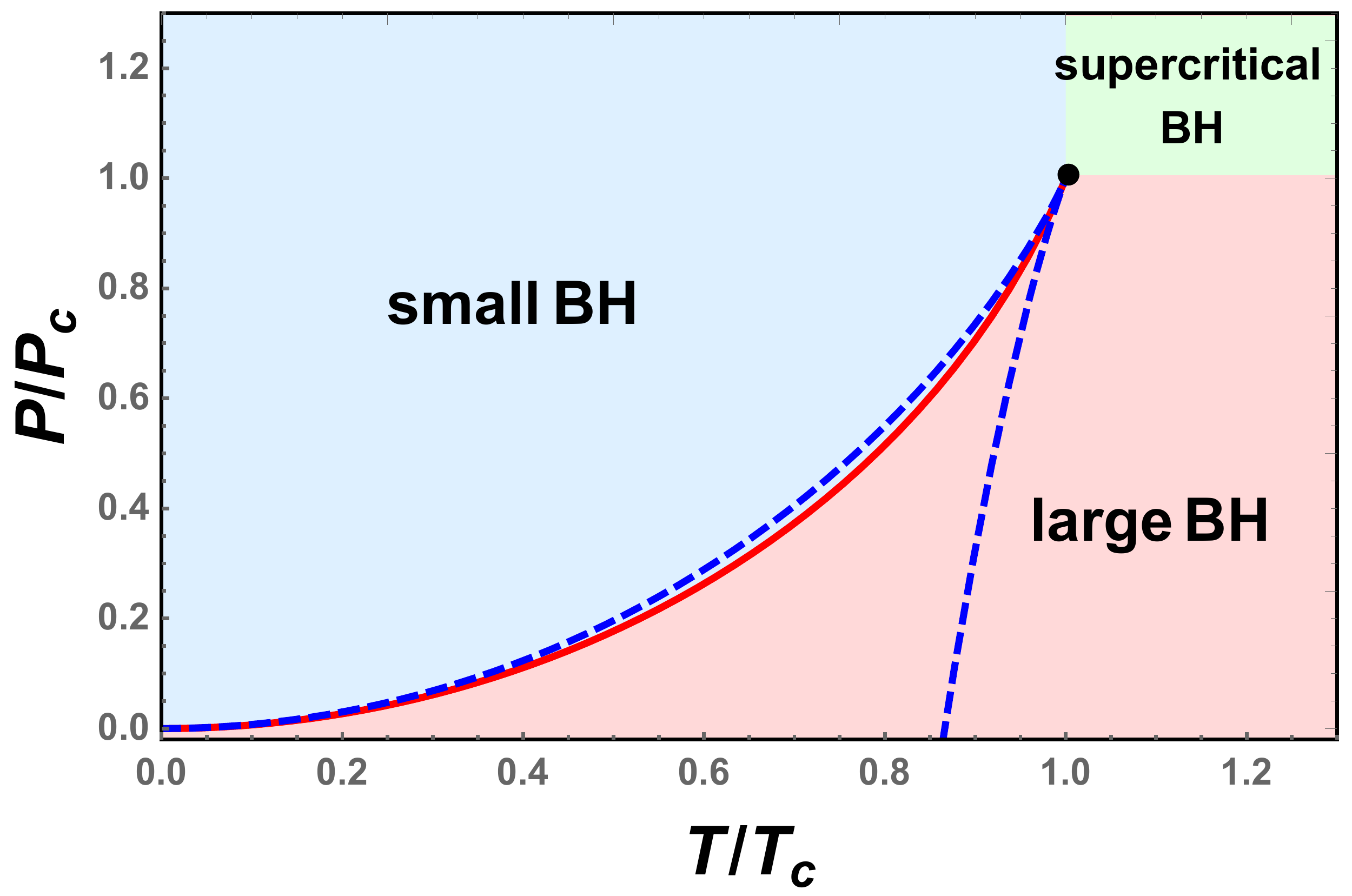}}
\subfigure[]{\label{TVphase}
\includegraphics[width=7cm]{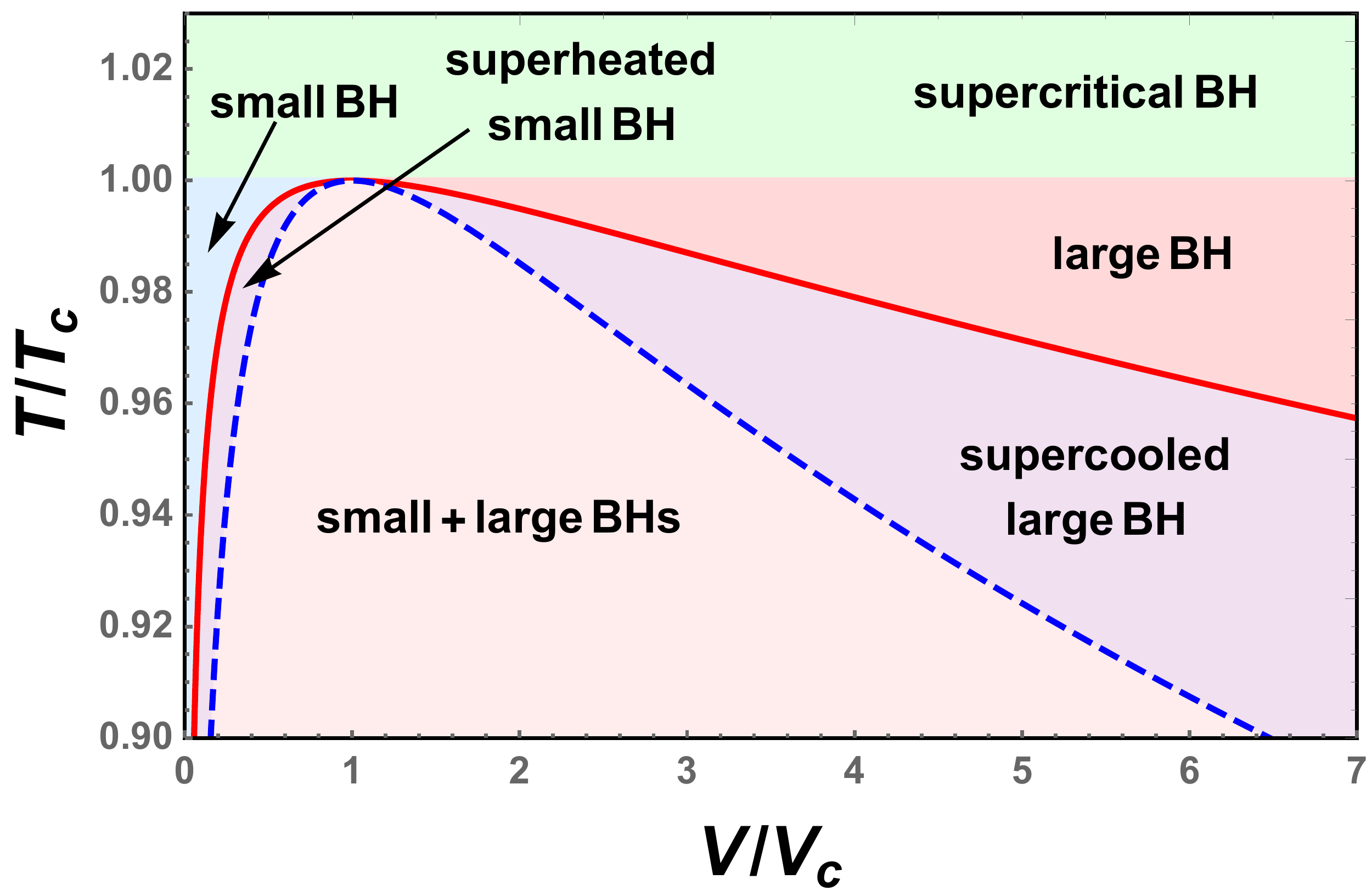}}}
\caption{Phase structures of the five-dimensional neutral GBA-dS black hole. The red solid curves are the coexistence curve of small and large black holes. The blue dashed curves are the spinodal curves. (a) $\tilde{P}$-$\tilde{T}$ diagram. (b) $\tilde{T}$-$\tilde{V}$ diagram.}\label{ppPTphase}
\end{figure}

The action describing a $d$-dimensional charged GB-AdS black hole is
\begin{eqnarray}
 &S=\int d^{d}x\sqrt{-g}
 \left(\frac{1}{16\pi G_{d}}(\mathcal{R}
    -2\Lambda
    +\alpha_{\text{GB}}\mathcal{L}_{\text{GB}})
  -\mathcal{L}_{\text{matter}}\right), \label{action}
\end{eqnarray}
where
\begin{eqnarray}
   \mathcal{L}_{\text{GB}}~~&=&\mathcal{R}_{\mu\nu\gamma\delta}
   \mathcal{R}^{\mu\nu\gamma\delta}
                    -4\mathcal{R}_{\mu\nu}\mathcal{R}^{\mu\nu}+\mathcal{R}^{2},\\
 \mathcal{L}_{\text{matter}}&=&4\pi \mathcal{F}_{\mu\nu}\mathcal{F}^{\mu\nu},
\end{eqnarray}
and $\alpha_{\text{GB}}$ is the GB coupling constant. The Maxwell field strength is defined as
$\mathcal{F}_{\mu\nu}=\partial_{\mu}\mathcal{A}_{\nu}-\partial_{\nu}\mathcal{A}_{\mu}$
with $\mathcal{A}_{\mu}$ the vector potential. The metric of the black hole is
\begin{eqnarray}
 ds^{2}=-f(r)dt^{2}+f^{-1}(r)dr^{2}+r^{2}(d\theta^{2}+\sin^{2}\theta d\phi^{2}+\cos^{2}\theta d\Omega^{2}_{d-4}),\label{metric}
\end{eqnarray}
with the metric function given by \cite{Boulware,Cai2,Wiltshire,Cvetic2}
\begin{eqnarray}
 f(r)=1+\frac{r^{2}}{2\alpha}
   \left(1-\sqrt{1 + \frac{2\alpha}{d-2} \left(\frac{32\pi M}{\Sigma_{d-2} r^{d-1}}-\frac{4 Q^{2}}{(d-3)r^{2d-4}}+\frac{4\Lambda}{d-1}\right)}\right),\label{metrf}
\end{eqnarray}
where $\alpha=(d-3)(d-4)\alpha_{\text{GB}}$ and $\Sigma_{d-2}$ is the area of a unit $(d-2)$-dimensional sphere. The parameters $M$ and $Q$ are the black hole mass and charge, respectively. In the extended phase space, the cosmological constant $\Lambda$ was interpreted as the thermodynamic pressure \cite{Kastor}
\begin{eqnarray}
 P=-\frac{1}{8\pi}\Lambda.
\end{eqnarray}
It was shown that the five-dimensional neutral GB-AdS black hole demonstrates a small-large black hole phase transition. Especially, the coexistence curve of the small and large black holes has an analytical expression, which allows us to exactly study the thermodynamics and phase transition. Moreover, it also provides us an opportunity to determine the effect of the GB coupling constant on the black hole thermodynamics. Therefore, in the following, we will take $d$=5 and $Q$=0. Then the metric function (\ref{metrf}) will be of the following form
\begin{eqnarray}
 f(r)=1+\frac{r^{2}}{2\alpha}
   \left(1-\sqrt{1 + \frac{32M\alpha}{3\pi r^{4}}-\frac{16P\pi\alpha}{3}}\right).
\end{eqnarray}
Solving the above equation, we obtain the black hole mass
\begin{eqnarray}
 M=\frac{\pi}{8}(4P\pi r_{\rm h}^{4}+3r_{\rm h}^{2}+3\alpha).
\end{eqnarray}
According to Ref.~\cite{Cai}, the black hole mass should be treated as the enthalpy $H\equiv M$. Other thermodynamic quantities read
\begin{eqnarray}
 T&=&\frac{8\pi P r_{\rm h}^{3}+3r_{\rm h}}{6\pi r_{\rm h}^{2}+12\pi\alpha},\quad
 S=\frac{\pi^{2}r_{\rm h}}{2}(r_{\rm h}^{2}+6\alpha),  \label{EntropyGBBH}\\
 \mathcal{A}&=&-\frac{\pi}{8}\frac{32\pi P r_{\rm h}^{4}+9r_{\rm h}^{2}-6\alpha}{r_{\rm h}^{2}+2\alpha},\quad
 V=\frac{\pi^{2}r_{\rm h}^{4}}{2},\quad
 v=\frac{4}{3}r_{\rm h}.
\end{eqnarray}
Here $\mathcal{A}$ is a conjugate quantity to the GB coefficient $\alpha$. The parameters $V$ and $v$ are, respectively, the thermodynamic volume and specific volume of the black hole system. It is easy to check that the following first law and Smarr relation hold
\begin{eqnarray}
 dH&=&TdS+VdP+\mathcal{A}d\alpha,\\
 2H&=&3TS-2PV+2\mathcal{A}\alpha.
\end{eqnarray}
The state equation for this black hole system is
\begin{eqnarray}
 P=\frac{T}{v}-\frac{2}{3\pi v^{2}}+\frac{32T\alpha}{9v^{3}}.
\end{eqnarray}
Obviously, the state equation closely depends on the GB coupling $\alpha$. For non-vanishing $\alpha$, there exists a small-large black hole phase transition of VdW type. The critical point can be obtained by solving $(\partial_{v}P)_{T}=(\partial_{v,v}P)_{T}=0$, which gives \cite{Mol}
\begin{eqnarray}
 P_{\rm c}=\frac{1}{48\pi\alpha},\quad
 T_{\rm c}=\frac{1}{2\pi\sqrt{6\alpha}},\quad
 V_{\rm c}=18\pi^{2}\alpha^{2},\quad
 v_{\rm c}=\sqrt{\frac{32\alpha}{3}}.
\end{eqnarray}
In the reduced parameter space, the state equation can be expressed as
\begin{eqnarray}
 \tilde{P}=\frac{3\tilde{T}}{\tilde{v}}-\frac{3}{\tilde{v}^{2}}
 +\frac{\tilde{T}}{\tilde{v}^{3}}, \label{resteq}
\end{eqnarray}
where the reduced quantities are defined by
\begin{eqnarray}
 \tilde{P}=\frac{P}{P_{\rm c}},\quad
 \tilde{T}=\frac{T}{T_{\rm c}},\quad
 \tilde{v}=\frac{v}{v_{\rm c}}.
\end{eqnarray}
It is quite clear that the reduced state equation (\ref{resteq}) is independent of the parameter $\alpha$, which is mainly because that the neutral GB-AdS black hole is a single-characteristic-parameter system \cite{Wei6}. Moreover, we can express the state function with the thermodynamic volume
\begin{eqnarray}
 P&=&\frac{3\pi^{3/2}\alpha T}{2\times(2V)^{3/4}}
 +\frac{3\sqrt{\pi}T}{4\times(2V)^{1/4}}
 -\frac{3}{8\sqrt{2V}},\label{pvsta}
\end{eqnarray}
or in the reduced parameter space
\begin{eqnarray}
 \tilde{P}&=&\frac{\tilde{T}}{\tilde{V}^{3/4}}
 -\frac{3}{\sqrt{\tilde{V}}}
 +\frac{3\tilde{T}}{\tilde{V}^{1/4}}.\label{rpvsta}
\end{eqnarray}
Now, let focus on the spinodal curve first, which satisfies
\begin{eqnarray}
 (\partial_{\tilde{V}}\tilde{P})_{\tilde{T}}=0.
\end{eqnarray}
This curve divides the metastable phase from the unstable phase, and the heat capacity diverges along this curve. Solving the equation, one can obtain the spinodal curve. Here we show a compact form of the spinodal curve
\begin{eqnarray}
 \tilde{T}_{\rm sp}=\frac{2\tilde{V}^{1/4}}{1+\sqrt{\tilde{V}}}.
\end{eqnarray}
In the reduced parameter space, the analytical coexistence curve is \cite{Mol}
\begin{eqnarray}
 \tilde{P}=\frac{1}{2}(3-\sqrt{9-8\tilde{T}^{2}}),
\end{eqnarray}
or
\begin{eqnarray}
 \tilde{T}=\sqrt{\frac{\tilde{P}(3-\tilde{P})}{2}}.
\end{eqnarray}
Interestingly, we obtain the formula of the coexistence curve in the $\tilde{T}$-$\tilde{V}$ diagram, which reads
\begin{eqnarray}
 \tilde{T}=\frac{3\tilde{V}^{1/4}(1+\sqrt{\tilde{V}})}{1+4\sqrt{\tilde{V}}+\tilde{V}}.
\end{eqnarray}
With these results, we show the phase structures in Fig. \ref{ppPTphase}. From Fig. \ref{PTphase}, we show three black hole phases, the small black hole, large black hole, and the supercritical black hole. The red solid curve is the coexistence curve of the small and large black holes. It starts from the origin, then increases with the temperature, and ends at the critical point marked with a black dot. Top and bottom blue dashed curves are the spinodal curves for the large and small black holes, respectively. Note that the small black hole spinodal curve starts at $\tilde{T}=\sqrt{3}/2$. The phase structure is also shown in the $\tilde{T}$-$\tilde{V}$ diagram in Fig. \ref{TVphase}. Two more metastable phases, the superheated small black hole phase and the supercooled large black hole phase, are clearly presented.

Next, we would like to examine the change of the thermodynamic volume among the black hole phase transition. After a simple calculation, we obtain the thermodynamic volumes $\tilde{V}_{\rm s}$ and $\tilde{V}_{\rm l}$ for the small and large black holes in terms of the phase transition temperature
\begin{eqnarray}
 \tilde{V}_{\rm s}&=&\left(\frac{2\tilde{T}-\sqrt{6}\sqrt{2\tilde{T}^{2}-3+\sqrt{9-8\tilde{T}^{2}}}}
   {3-\sqrt{9-8\tilde{T}^{2}}}\right)^{4},\label{vvs}\\
 \tilde{V}_{\rm l}&=&\left(\frac{2\tilde{T}+\sqrt{6}\sqrt{2\tilde{T}^{2}-3+\sqrt{9-8\tilde{T}^{2}}}}
   {3-\sqrt{9-8\tilde{T}^{2}}}\right)^{4}.\label{vvl}
\end{eqnarray}
In terms of the phase transition pressure, these volumes can also be expressed as
\begin{eqnarray}
 \tilde{V}_{\rm s}&=&\frac{(\sqrt{3-3\tilde{P}}-\sqrt{3-\tilde{P}})^{4}}{4\tilde{P}^{2}},\\
 \tilde{V}_{\rm l}&=&\frac{(\sqrt{3-3\tilde{P}}+\sqrt{3-\tilde{P}})^{4}}{4\tilde{P}^{2}}.
\end{eqnarray}
Moreover, we find the follow interesting relations
\begin{eqnarray}
 \tilde{V}_{\rm l}+\tilde{V}_{\rm s}=14+\frac{36-48\tilde{P}}{\tilde{P}^{2}},\quad
 \tilde{V}_{\rm l}*\tilde{V}_{\rm s}=1.
\end{eqnarray}

\begin{figure}
\center{\subfigure[]{\label{PDVCT}
\includegraphics[width=7cm]{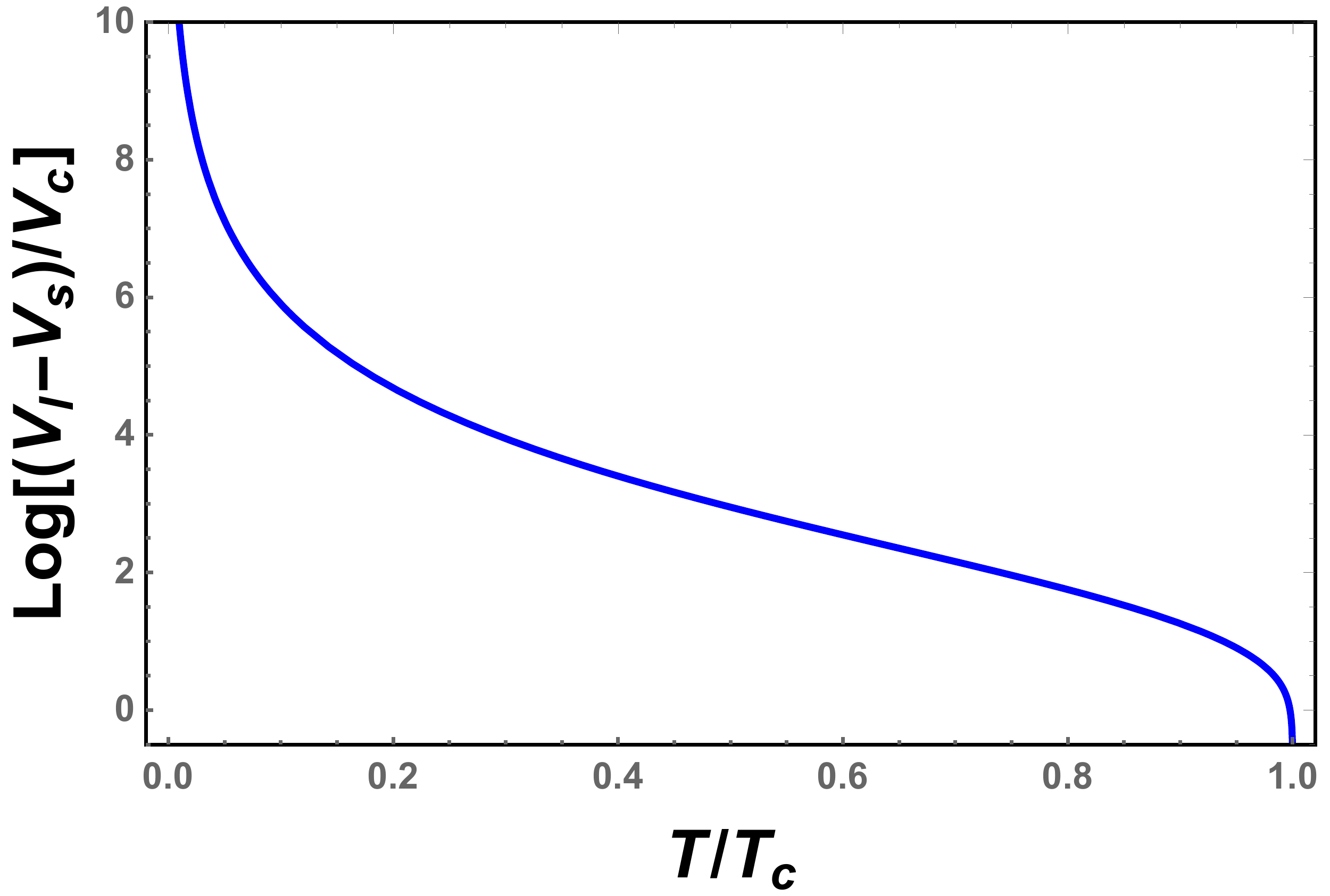}}
\subfigure[]{\label{TDVCP}
\includegraphics[width=7cm]{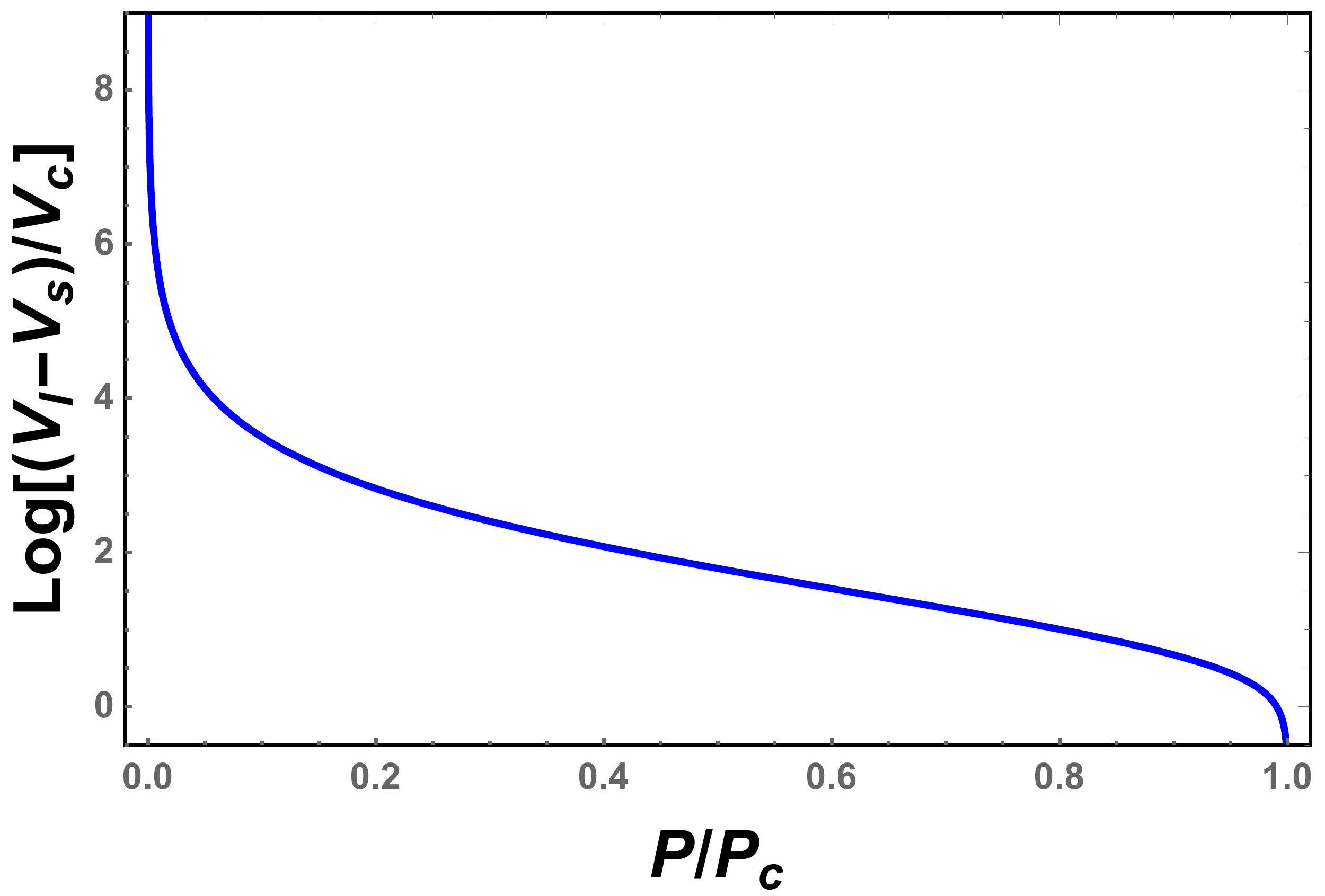}}}
\caption{Change of the volumes $\Delta\tilde{V}=\tilde{V}_{\rm l}-\tilde{V}_{\rm s}$ as a function of the phase transition temperature and pressure. (a) $\log_{10}\Delta\tilde{V}$ vs $\tilde{T}$. (b) $\log_{10}\Delta\tilde{V}$ vs $\tilde{P}$.}\label{ppDVCP}
\end{figure}

We describe the change of the thermodynamic volume $\Delta\tilde{V}=\tilde{V}_{\rm l}-\tilde{V}_{\rm s}$ in Fig. \ref{ppDVCP} as a function of the phase transition temperature and pressure, respectively. From the figures, we find that with the increase of the temperature or the pressure, $\Delta\tilde{V}$ decreases, and when the critical point is approached, $\Delta\tilde{V}$ vanishes. Note that at the critical point, $\tilde{V}_{\rm l}=\tilde{V}_{\rm s}$, and thus $\log_{10}\Delta\tilde{V}=-\infty$. We have made a cut-off in the figures at the critical point.

Since the thermodynamic volume has analytical form, we can expand the change of the volume $\Delta\tilde{V}$ near the critical point. The results are
\begin{eqnarray}
 \Delta\tilde{V}&=&8\sqrt{6}(1-\tilde{T})^{\frac{1}{2}}
 +118\sqrt{6}(1-\tilde{T})^{\frac{3}{2}}
  +\mathcal{O}(1-\tilde{T})^{\frac{5}{2}},\\
 \Delta\tilde{V}&=&4\sqrt{6}(1-\tilde{P})^{\frac{1}{2}}
 +17\sqrt{6}(1-\tilde{P})^{\frac{3}{2}}
  +\mathcal{O}(1-\tilde{P})^{\frac{5}{2}}.
\end{eqnarray}
Therefore, at the critical point, $\Delta\tilde{V}$ has a universal exponent $\frac{1}{2}$. Considering the behavior of the $\Delta\tilde{V}$, it can serve as an order parameter to describe the small-large black hole phase transition. The parametric study of the phase transition for the GB-AdS black hole can also be found in Ref. \cite{MiaoXu}.

\section{Ruppeiner geometry and scalar curvature}
\label{rgsc}

In this section, we would like to study the Ruppeiner geometry for the five-dimensional neutral GB-AdS black hole. Employing the scalar curvature of the geometry, we can examine the microstructures of the black hole.

\begin{figure}
\center{\subfigure[]{\label{PRnT04}
\includegraphics[width=7cm]{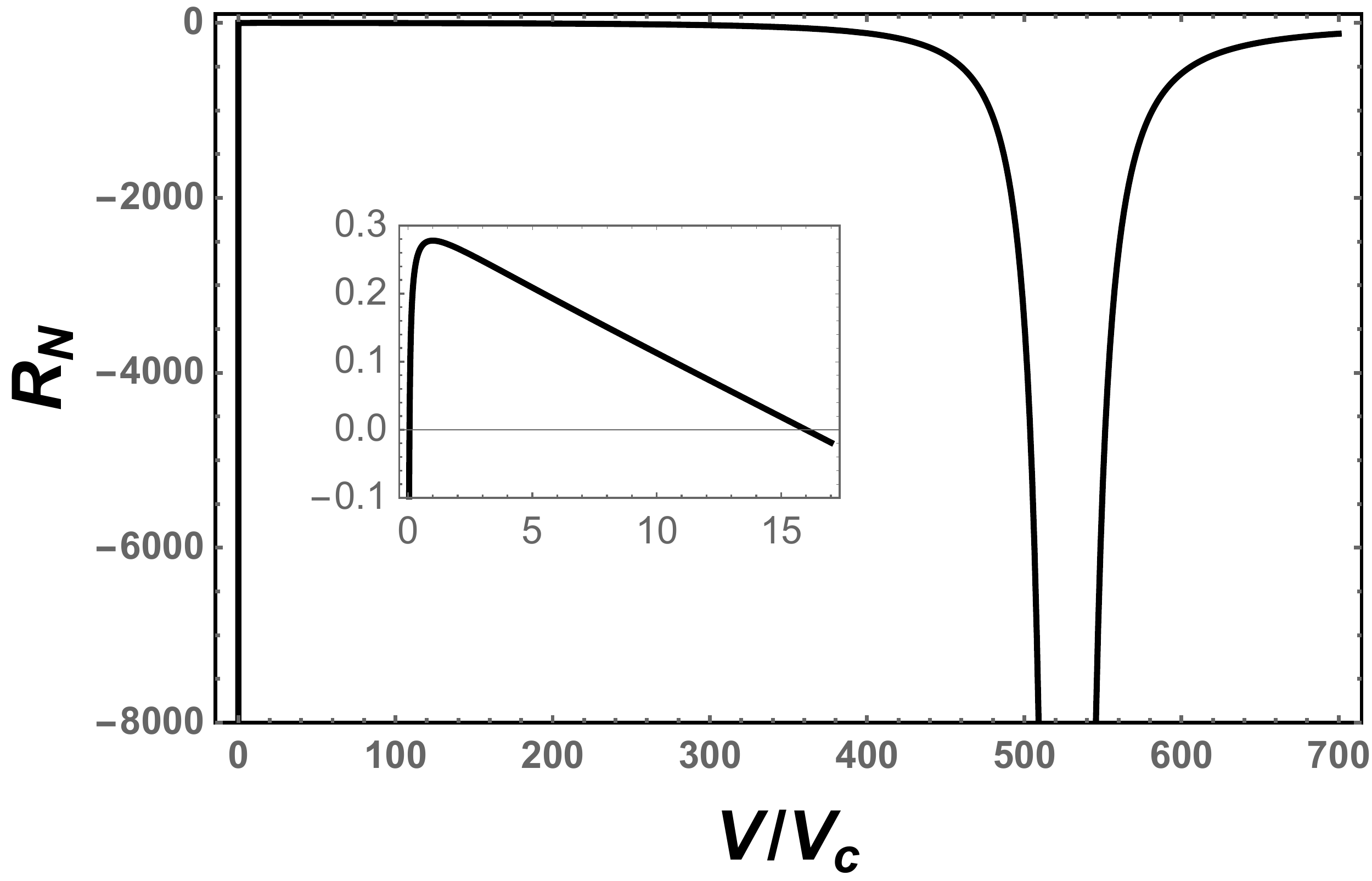}}
\subfigure[]{\label{TRnT08}
\includegraphics[width=7cm]{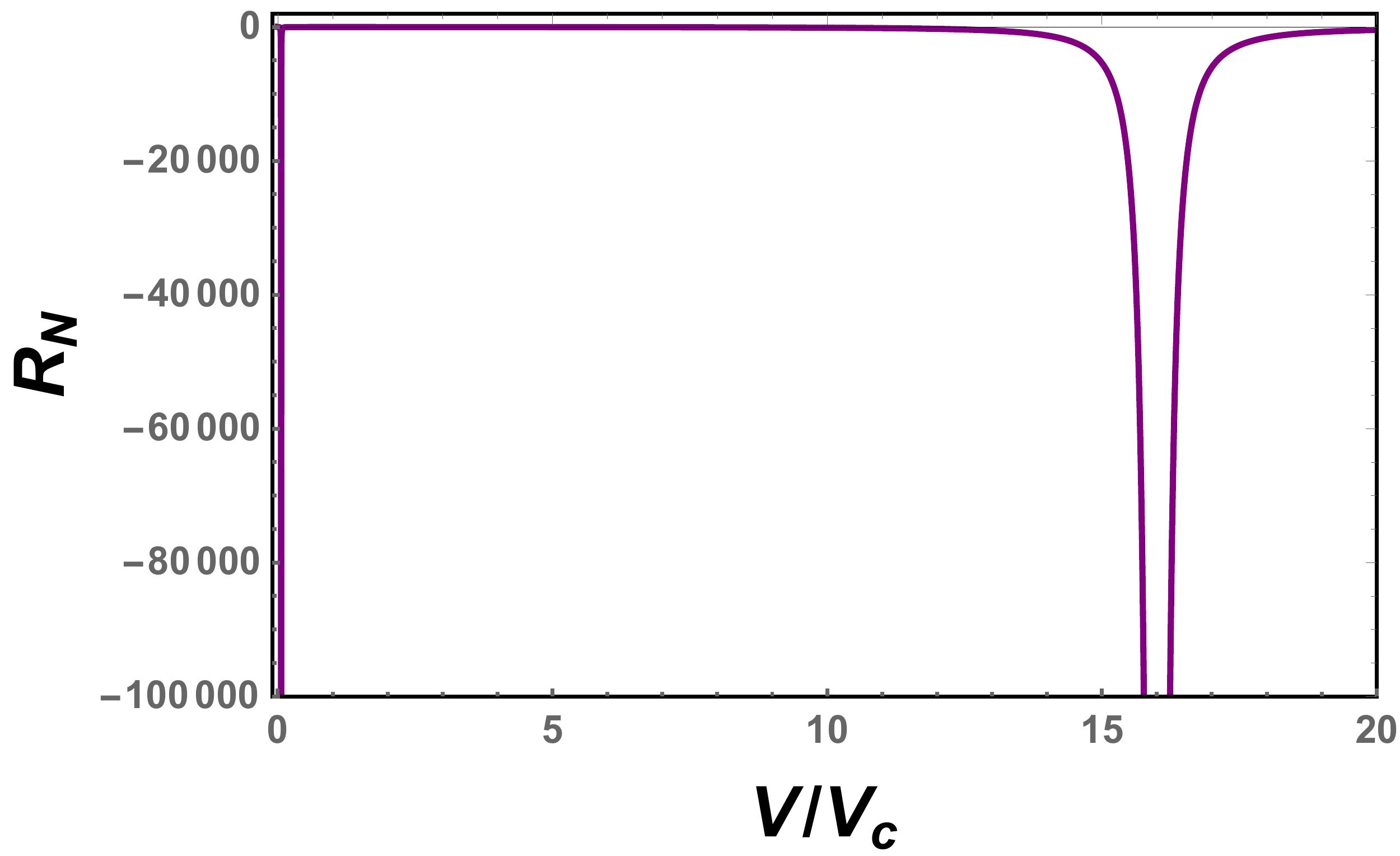}}
\subfigure[]{\label{TRnT09}
\includegraphics[width=7cm]{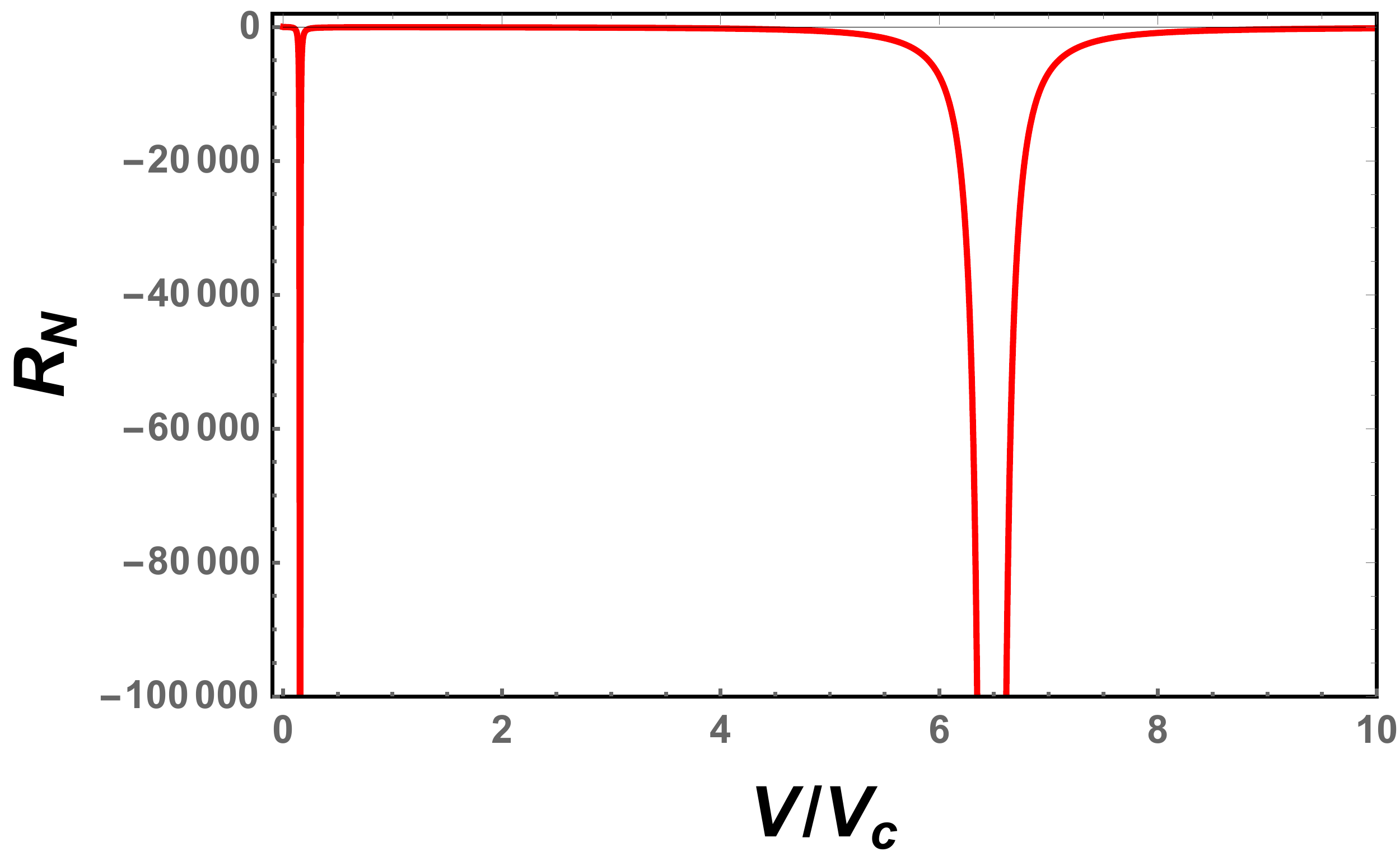}}
\subfigure[]{\label{TRnT10}
\includegraphics[width=7cm]{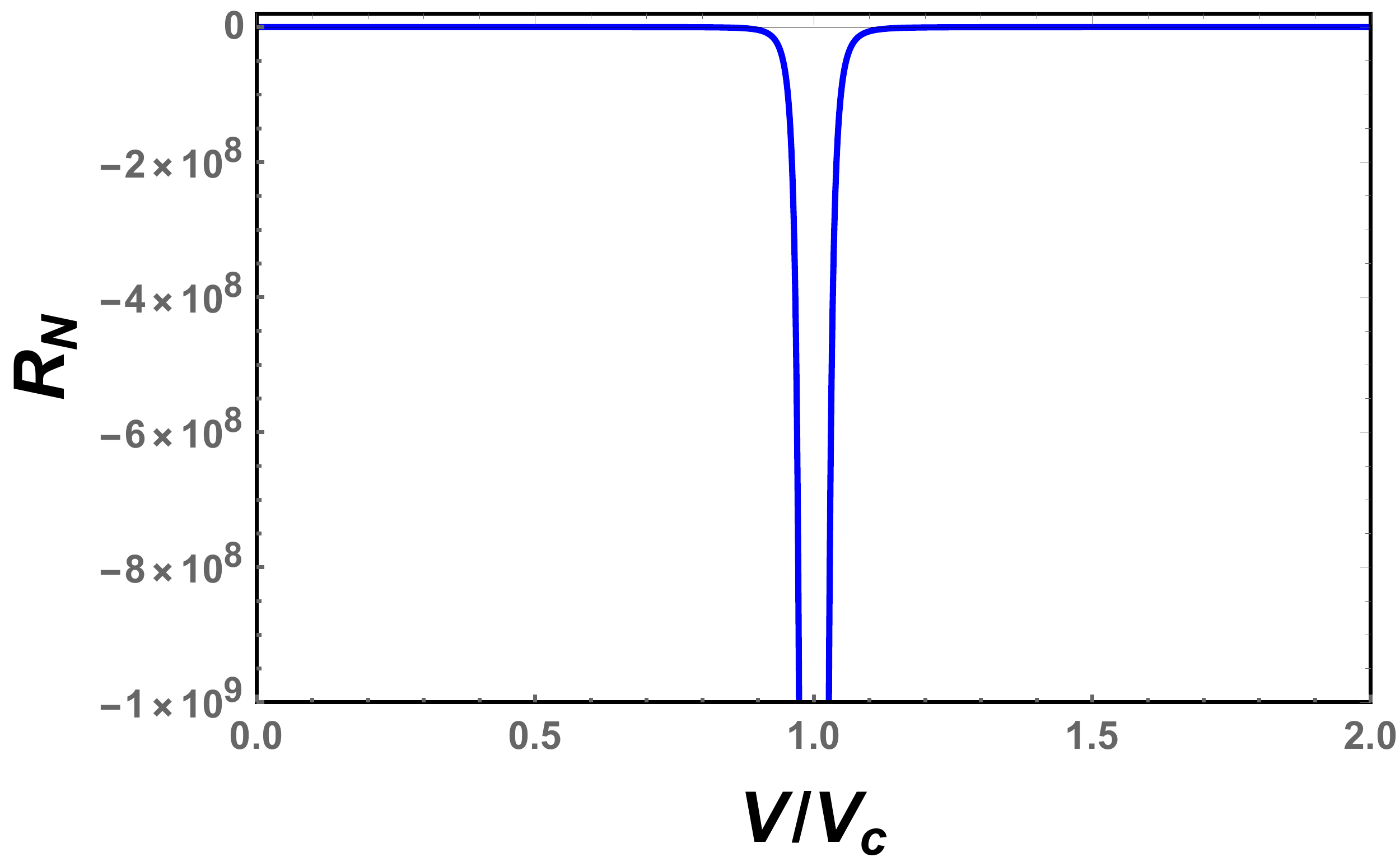}}}
\caption{Behavior of the normalized scalar curvature $R_{\rm N}$ as a function of the reduced volume $\tilde{V}$. (a) $\tilde{T}$=0.4. (b) $\tilde{T}$=0.8. (c) $\tilde{T}$=0.9. (d) $\tilde{T}$=1.0.}\label{ppRnT04}
\end{figure}

By using Eqs. (\ref{xxy}) and (\ref{pvsta}), the line element of the Ruppeiner geometry reads
\begin{eqnarray}
 dl^{2}=\frac{C_{V}}{T^{2}}dT^{2}+\frac{3\left(6\pi^{3/2}T\alpha+\sqrt{2\pi V }T-(2V)^{1/4}\right)}{16\times 2^{3/4}V^{7/4}T}dV^{2}.
\end{eqnarray}
For this neutral GB-AdS black hole, the heat capacity at fixed volume and GB coefficient vanishes. Therefore, we would like to adopt the normalized scalar curvature $R_{\rm N}=C_{V}*R$ defined in \cite{LiuLiu}, which is found to be a good indicator of the microscopic interaction of the black holes and has a critical phenomena as we expected. After a simple calculation, the normalized scalar curvature is
\begin{eqnarray}
 R_{\rm N}=\frac{2^{1/4}\sqrt{V}-2\sqrt{2\pi}TV^{3/4}-12\pi^{3/2}\alpha TV^{1/4}}{2^{3/4}\left(6\pi^{3/2}\alpha T+\sqrt{2\pi V}T-(2V)^{1/4}\right)}.
\end{eqnarray}
Obviously, $R$ closely depends on the coupling constant $\alpha$. In the reduced parameter space, it is
\begin{eqnarray}
 R_{\rm N}=\frac{2\tilde{V}^{1/4}\left(\tilde{V}^{1/4}
 -\tilde{T}\left(1+\sqrt{\tilde{V}}\right)\right)}
 {\left(\tilde{T}-2\tilde{V}^{1/4}+\tilde{T}\sqrt{\tilde{V}}\right)^{2}}.\label{norscu}
\end{eqnarray}
Similar to the charged AdS black hole \cite{WeiWeiWei}, this normalized scalar curvature $R_{\rm N}$ is independent of $\alpha$. We show the behavior of $R_{\rm N}$ in Fig. \ref{ppRnT04} for $\tilde{T}$=0.4, 0.8, 0.9, and 1.0. When $\tilde{T}$=0.4, 0.8 and 0.9, we observe two negative divergent points. And these two points get closer with the increase of $\tilde{T}$. At the critical temperature $\tilde{T}$=1.0, these two divergent points coincide with each other at $\tilde{V}$=1. More importantly, for $\tilde{T}$=0.4, we observe a positive $R_{\rm N}$ near the region $\tilde{V}\in(0.06, 16)$, which, according to the interpretation of the Ruppeiner geometry, implies a repulsive interaction among the microstructure of the black hole system.

From the expression of normalized scalar curvature (\ref{norscu}), we find that $R_{\rm N}$ exactly diverges at the spinodal curve. Moreover, solving $R_{\rm N}=0$, we can obtain the sign-changing curve, which is
\begin{eqnarray}
 \tilde{T}_{0}=\frac{\tilde{T}_{\rm sp}}{2}=\frac{\tilde{V}^{1/4}}{1+\sqrt{\tilde{V}}}.
\end{eqnarray}
Therefore, the temperature $T_{0}$ is half of that of the spinodal curve, which is similar to the charged AdS black hole \cite{WeiWeiWei}.

\begin{figure}
\center{
\includegraphics[width=8cm]{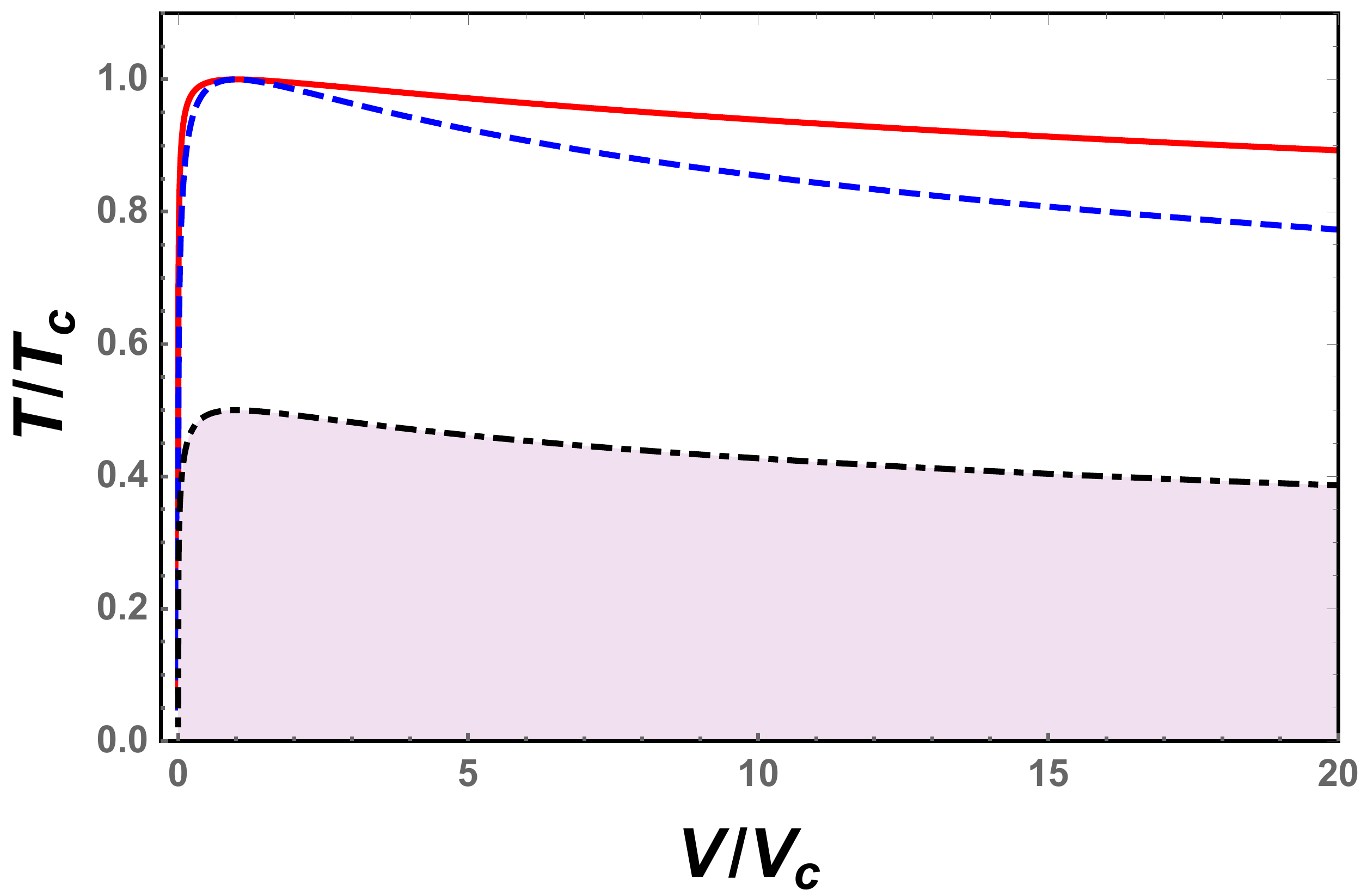}}
\caption{Behaviors of the coexistence curve (red solid line), spinodal curve (blue dashed line), and sign-changing curve (black dot dashed line) in the $\tilde{T}$-$\tilde{V}$ diagram. The region marked in light purple color has positive $R_{\rm N}$, otherwise, it has negative $R_{\rm N}$.}\label{ppCurve}
\end{figure}

In Fig. \ref{ppCurve}, we list the coexistence curve (red solid line), spinodal curve (blue dashed line), and the sign-changing curve (black dot dashed line) in the $\tilde{T}$-$\tilde{V}$ diagram. The region marked in light purple color below the sign-changing curve has positive $R_{\rm N}$, while the region above it has negative $R_{\rm N}$. It is clear that the spinodal curve and the sign-changing curve are both under the coexistence curve. Consider that the state equation (\ref{rpvsta}) does not hold in the coexistence region of small and large black holes, the region of positive $R_{\rm N}$ will be excluded, and thus for the five-dimensional neutral GB-AdS black hole system, $R_{\rm N}$ is always negative, which implies that only attractive interaction exists among the microstructures. This property is quite different from that of the charged AdS black hole shown in Refs. \cite{LiuLiu,WeiWeiWei}, where there can exist a repulsive interaction for the small black holes. So not all the black hole systems allow the existence of the repulsive interaction.

\begin{figure}
\center{
\includegraphics[width=8cm]{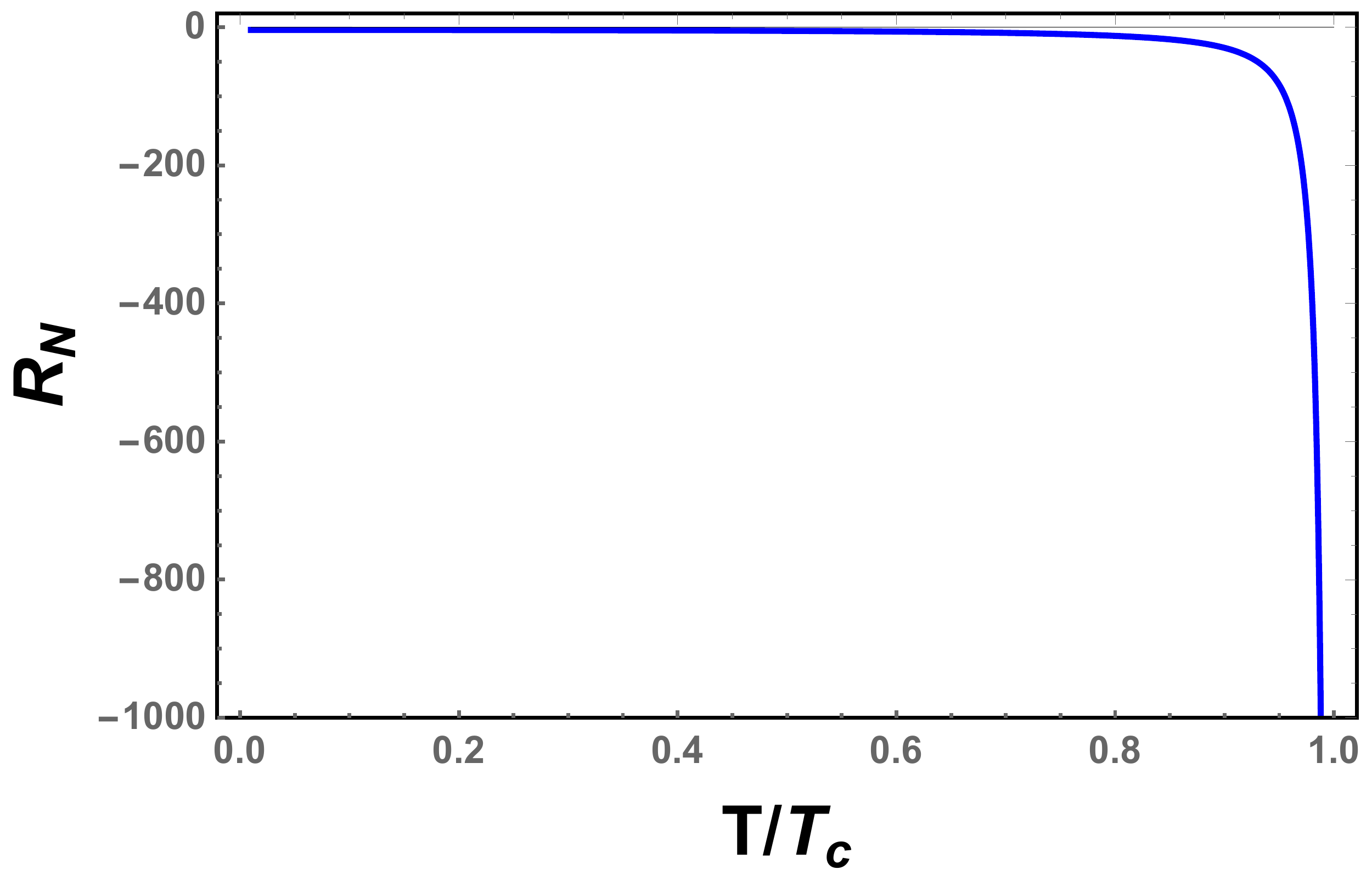}}
\caption{Behavior of $R_{\rm N}$ along the coexistence curve of the small and large black holes.}\label{ppRNCSL}
\end{figure}

Next, we would like to examine the behavior of $R_{\rm N}$ along the coexistence curve and near the critical point.

With the help of (\ref{vvs}) and (\ref{vvl}), we express $R_{\rm N}$ along the saturated small and large black hole curves. The result strongly suggests that the normalized scalar curvature $R_{\rm N}$ along these two curves shares the same expression, which is
\begin{eqnarray}
 R_{\rm N}=\frac{6\tilde{T}^{2}-(3-2\tilde{T}^{2})\sqrt{9-8\tilde{T}^{2}}-7}
 {4\left(1-\tilde{T}^{2}\right)^{2}}.
\end{eqnarray}
This result is novel, and indicates that the interaction among the microstructures is the same before and after the phase transition. However the microstructure does change. Therefore, it gives us the first example that the black hole microscopic interaction does not change during the phase transition. Moreover, the numerator of $R_{\rm N}$ is negative for $\tilde{T}\in(0, 1)$, which leads to a negative $R_{\rm N}$. We also plot it in Fig. \ref{ppRNCSL}. From it, we observe a negative $R_{\rm N}$. With the increase of $\tilde{T}$, $R_{\rm N}$ decreases and diverges at the critical temperature.

Near the critical point, we expand $R_{\rm N}$ as
\begin{eqnarray}
 R_{\rm N}=-\frac{1}{8}(1-\tilde{T})^{-2}-\frac{13}{8}(1-\tilde{T})^{-1}-\frac{27}{32}
 +\mathcal{O}(1-\tilde{T}).
\end{eqnarray}
This implies that $R_{\rm N}$ has a universal exponent $2$ near the critical point. Moreover, ignoring the high orders, we obtain the following relation
\begin{eqnarray}
 R_{\rm N}(1-\tilde{T})^{2}=-\frac{1}{8},
\end{eqnarray}
which is the same constant as the VdW fluid and four-dimensional charged AdS black hole, while it is different from that of the higher-dimensional charged AdS black holes \cite{WeiWeiWei}.

\section{Conclusions and discussions}
\label{Conclusion}

In this paper, we analytically studied the thermodynamic phase transition and Ruppeiner geometry for the five-dimensional neutral GB-AdS black hole. Combining with the phase transition, we understood the microscopic properties of the black hole by analyzing the behavior of the scalar curvature of the Ruppeiner geometry along the coexistence curve and near the critical point.

By using analytical form of the coexistence curve of the small and large black holes, we showed the phase structure of the black hole in the $\tilde{P}$-$\tilde{T}$ and $\tilde{T}$-$\tilde{V}$ diagrams, respectively. Especially, in the $\tilde{T}$-$\tilde{V}$ diagram, two metastable black hole phases, the superheated small black hole phase and supercooled large black hole phase, were displayed. The spinodal curves of the small and large black holes were also shown in the phase diagrams, along which the heat capacity diverges. We also obtained the thermodynamic volumes along the saturated small and large black holes. Their change $\Delta\tilde{V}$ decreases with the phase transition temperature or pressure, and vanishes at the critical point. The result also shows that $\Delta\tilde{V}$ has a universal exponent $\frac{1}{2}$ at the critical point. Therefore, $\Delta\tilde{V}$ can serve as an order parameter to describe the small-large black hole phase transition.

\begin{table}[h]
\begin{center}
\begin{tabular}{ccccccc}
  \hline\hline
 & saturated SBH/liquid & saturated LBH/gas & critical point & critical exponent & $R_{\rm N}(1-\tilde{T})^{2}$\\\hline
 VdW fluid & negative & negative & negative divergence & 2 & -$\frac{1}{8}$ \cite{WeiWeiWei}\\
 $d=4$ charged AdS BH & negative~$\&$~positive & negative & negative divergence & 2 & -$\frac{1}{8}$ \cite{LiuLiu}\\
 $d\geq5$ charged AdS BH & negative~$\&$~positive & negative & negative divergence & 2 & $<$-$\frac{1}{8}$ \cite{WeiWeiWei}\\
  $d=5$ neutral GB-AdS BH & negative & negative & negative divergence & 2 & -$\frac{1}{8}$\\\hline\hline
\end{tabular}
\caption{Properties of the normalized scalar curvature $R_{\rm N}$ along the saturated small black hole (liquid) and large black hole (gas) curves, and near the critical point for the VdW fluid, $d$-dimensional charged AdS black hole, and five-dimensional neutral GB-AdS black hole.}\label{tab1}
\end{center}
\end{table}

Then we constructed the Ruppeiner geometry and calculated the scalar curvature. Following Ref. \cite{LiuLiu}, we adopted the normalized scalar curvature $R_{\rm N}$ to test the microstructure of the five-dimensional neutral GB-AdS black hole. In order to give a clear comparison of the properties of $R_{N}$ for the VdW fluid, charged AdS black hole, and five-dimensional neutral GB-AdS black hole, we summarize the results in Table \ref{tab1}. From the table, we find some results of $R_{\rm N}$ for the five-dimensional neutral GB-AdS BH: i) $R_{N}$ is negative, which indicates attractive interaction of the microstructure of the thermodynamic systems. ii) At the critical point, $R_{N}$ goes to negative infinity. iii) $R_{N}$ has a critical exponent 2 and $R_{\rm N}(1-\tilde{T})^{2}=-\frac{1}{8}$. One more intriguing property we observed for the neutral GB-AdS black hole is that, the normalized scalar curvature $R_{N}$ shares the same form for both the saturated small and large black holes. As we know, when the black hole system crosses the first order phase transition, its microstructures will experience a significant change. However, from the result of the normalized scalar curvature $R_{N}$, these two different black hole systems with different microstructures have the same interaction. This reveals a particular interesting nature for the AdS black hole in GB gravity.

In summary, we analytically investigated the property of the microstructure for the neutral AdS black hole in GB gravity, by combining with the thermodynamic phase transition and Ruppeiner geometry. The results show that the attractive interaction is dominant for the black hole system. Of particular interest is that during the small-large black hole phase transition, the scalar curvature implies that the interaction keeps unchanged while the microstructure of the black hole system does change. This is an interesting and important result for the black hole in the GB gravity. Our study is also worthwhile generalizing to charged and higher-dimensional AdS black hole in GB gravity. However, there may be no analytic result, and one needs to use the numerical calculation. Nevertheless, these will strengthen our knowledge on the black hole microstructures.

\section{Appendix: Line element of Ruppeiner geometry}

Let us start with the first law of the neutral GB-AdS black hole,
\begin{eqnarray}
 dU=TdS+\mathcal{A} d\alpha-PdV.
\end{eqnarray}
Here we consider the constant $\alpha$ case, thus
\begin{eqnarray}
 dU=TdS-PdV.\label{fflaw}
\end{eqnarray}
Deforming it, we have
\begin{eqnarray}
 dS=\frac{1}{T}dU+\frac{P}{T}dV.
\end{eqnarray}
Making use it, we can obtain the line element of the Ruppeiner geometry
\begin{eqnarray}
 dl^{2}&=&-\left(d\left(\frac{1}{T}\right)dU+d\left(\frac{P}{T}\right)dV\right)\nonumber\\
       &=&\frac{1}{T^{2}}dTdU+\frac{P}{T^{2}}dTdV-\frac{1}{T}dPdV.
\end{eqnarray}
Plunging the first law (\ref{fflaw}) into above equation, we will obtain
\begin{eqnarray}
 dl^{2}=\frac{1}{T}dSdT-\frac{1}{T}dPdV.
\end{eqnarray}
Here we take the temperature $T$ and the volume $V$ as the variables, so we expand the entropy $S$ and the pressure $P$ as
\begin{eqnarray}
 dS&=&\left(\frac{\partial S}{\partial T}\right)_{V}dT+\left(\frac{\partial S}{\partial V}\right)_{T}dV\\
 dP&=&\left(\frac{\partial P}{\partial T}\right)_{V}dT+\left(\frac{\partial P}{\partial V}\right)_{T}dV.
\end{eqnarray}
Further, one has
\begin{eqnarray}
 dl^2=\frac{1}{T}\left(
     \left(\frac{\partial S}{\partial T}\right)_{V}dT^{2}
    +\left(\frac{\partial S}{\partial V}\right)_{T}dTdV
    -\left(\frac{\partial P}{\partial T}\right)_{V}dTdV
    -\left(\frac{\partial P}{\partial V}\right)_{T}dV^{2}\right).
\end{eqnarray}
For a thermodynamic system, the Helmholtz free energy is $F=U-TS$, and the The
differential law is
\begin{eqnarray}
 dF=-SdT-PdV. \label{flaww}
\end{eqnarray}
Using it, it is easy to prove the following relation holds
\begin{eqnarray}
 \left(\frac{\partial P}{\partial T}\right)_{V}
 =\left(\frac{\partial S}{\partial V}\right)_{T}.
\end{eqnarray}
Then the line element reduces to
\begin{eqnarray}
 dl^2=\frac{1}{T}\left(\left(\frac{\partial S}{\partial T}\right)_{V}dT^{2}
    -\left(\frac{\partial P}{\partial V}\right)_{T}dV^{2}\right).
\end{eqnarray}
It is clear that the corresponding metric become diagonal. Employing the differential law (\ref{flaww}), it easy to get $S=-\left(\frac{\partial F}{\partial T}\right)_{V}$ and $P=-\left(\frac{\partial F}{\partial V}\right)_{T}$. Therefore, the geometry can be expressed with the Helmholtz free energy,
\begin{eqnarray}
 dl^{2}=-\frac{1}{T}\left(\frac{\partial^{2}F}{\partial T^{2}}\right)_{V}d T^{2}
   +\frac{1}{T}\left(\frac{\partial^{2}F}{\partial V^{2}}\right)_{T}d V^{2}.
\end{eqnarray}

\section*{Acknowledgements}
This work was supported by the National Natural Science Foundation of China (Grants No. 11675064 and No. 11875151).

\end{document}